\documentstyle[12pt,worldsci,amsbsy,amssymb,epsfig]{article}
\newcounter{figs}
\begin{document}
\begin{flushright}
\begin{tabular}{r}
DFTT 41/95
\\
hep-ph/9507226
\\
July 1995
\end{tabular}
\end{flushright}
\title{SOME REMARKS ON
$ \boldsymbol{ \nu_\mu } \boldsymbol{ \leftrightarrows } \boldsymbol{ \nu_e } $
OSCILLATIONS OF TERRESTRIAL NEUTRINOS AND
ATMOSPHERIC NEUTRINO OSCILLATIONS\footnote{
Talk presented by S.M. Bilenky at the
{\em Workshop on Particle Theory and Phenomenology},
Iowa State University, Ames, Iowa,
May 1995.
}}
\author{
S.M. Bilenky$^{\mathrm{a,b}}$,
A. Bottino$^{\mathrm{a}}$,
C. Giunti$^{\mathrm{a}}$
and
C. W. Kim$^{\mathrm{c}}$
\\
\begin{tabular}{c} \em
$^{\mathrm{a}}$INFN, Sezione di Torino
and Dipartimento di Fisica Teorica, Universit\`a di Torino,
\\ \em
Via P. Giuria 1, 10125 Torino, Italy
\\ \em
$^{\mathrm{b}}$Joint Institute of Nuclear Research, Dubna, Russia
\\ \em
$^{\mathrm{c}}$Department of Physics and Astronomy,
The Johns Hopkins University,
\\ \em
Baltimore, Maryland 21218, USA.
\end{tabular}
}
\maketitle
\begin{center}
{ABSTRACT\vspace{-.5em}\vspace{0pt}}
\end{center}
\begin{quote}
We present the results
of an analysis of the data of reactor
and accelerator
neutrino oscillation experiments
in the framework of a model
with mixing of three massive neutrino fields
and
a neutrino mass hierarchy.
The results of the LSND experiment are commented.
Taking into account future long-baseline
neutrino oscillation experiments,
we have also analyzed the atmospheric neutrino data
in the framework of the model.
\end{quote}
\setlength{\baselineskip}{2.6ex}

The problem of neutrino masses and mixing is the central issue
in neutrino physics at present.
It is  generally believed that the experimental
investigation of this problem
could lead to a discovery of new physics beyond the standard model.
We will discuss here some possibilities to reveal neutrino mixing in
experiments with terrestrial neutrinos.

In accordance with
the hypothesis of neutrino mixing
\cite{PONTECORVO},
the flavor neutrino fields $\nu_{\ell L}$ are linear combinations
of the left-handed components of neutrino fields
$\nu_{iL}$
with masses $m_i$:
\begin{equation}
\nu_{\ell L}
=
\sum_{i}
U_{\ell i}
\nu_{iL}
\;,
\label{E201}
\end{equation}
where $U$ is a unitary mixing matrix.
{}From LEP experiments
it follows that
the number of flavor neutrino fields
$\nu_{\ell L}$ is equal to three.
We would like
to stress that LEP data do not provide
information about the number of massive neutrinos.
In fact,
if
$ m_i \ll m_Z $,
from Eq.(\ref{E201})
for the invisible width of the $Z$-boson
we have
\begin{equation}
\Gamma_{\mathrm{inv}}
=
\sum_{\ell,i}
\left| U_{\ell i} \right|^2
\Gamma_{0}(Z\to\nu\bar\nu)
\;,
\label{E202}
\end{equation}
where
$ \Gamma_{0}(Z\to\nu\bar\nu) $
is the decay width of the
$Z$-boson into a massless neutrino-antineutrino pair.
Independently of the number of neutrinos with
a definite mass,
from the unitarity of the mixing matrix we have
$ \displaystyle
\sum_{\ell,i} \left| U_{\ell i} \right|^2
=
n_{\ell}
$,
where $n_{\ell}$ is the number of flavor neutrinos.

The number of massive neutrinos depends on the neutrino mixing scheme.
If the total lepton number
$ L = L_e + L_\mu + L_\tau $
is conserved,
the massive neutrinos are Dirac particles
and the number of massive neutrinos is equal to three
(Dirac mass term).
If the total lepton number is not conserved,
the neutrinos with
definite masses are Majorana particles.
If the Lagrangian contains only left-handed flavor neutrino fields
$\nu_{\ell L}$,
then the number of massive Majorana neutrinos is equal to three
(Majorana mass term).
In the most general case of neutrino mixing
the mass term contains
both $\nu_{\ell L}$ and $\nu_{\ell R}$ fields
and the lepton number $L$ is not conserved
(Dirac and Majorana mass term).
In this case the mass matrix is
a $6\times6$ matrix,
there are six Majorana particles with definite mass
and
\begin{equation}
\nu_{\ell L}
=
\sum_{i=1}^{6}
U_{\ell i}
\nu_{iL}
\qquad
\mbox{and}
\qquad
\left( \nu_{\ell R} \right)^{c}
=
\sum_{i=1}^{6}
U_{\bar\ell i}
\nu_{iL}
\qquad
(\ell=e,\mu,\tau)
\;,
\label{E205}
\end{equation}
where
$ \displaystyle
\left( \nu_{\ell R} \right)^{c}
=
\cal{C} \overline{\nu}_{\ell R}^{T}
$
is the charge conjugated field.

The most popular mechanism of neutrino mass generation
is the see-saw mechanism
\cite{SEESAW}.
This mechanism is based on the assumption
that the violation of the lepton number is due to
the right-handed fields $\nu_{\ell R}$
and that the lepton number is violated at the
Grand Unification scale
$ M_{\mathrm{GUT}} $.
In this case in the spectrum
of the masses of Majorana
particles there are three light masses $m_i$ (neutrinos)
and three very heavy masses $M_i$
($ M_i \simeq M_{\mathrm{GUT}} $).
The neutrino masses are related to the heavy masses by
the relation
\begin{equation}
m_i
\simeq
{\displaystyle
\left( m_{\mathrm{F}}^{i} \right)^2
\over\displaystyle
M_i
}
\;,
\label{E206}
\end{equation}
where $m_{\mathrm{F}}^{i}$ is the mass of the charged lepton or up-quark
in the corresponding generation.

Thus,
in most models of neutrino mixing
the number of massive neutrino fields in the relation (\ref{E201})
is equal to three.
If the number of massive
Majorana neutrinos is more than three,
then,
due to the mixing (\ref{E205}),
active flavor neutrinos could be transformed into
sterile states.

Let us consider now the oscillations of terrestrial and atmospheric neutrinos
in the minimal case
of neutrino mixing with three massive neutrino fields.
We will enumerate
the neutrino masses as follows:
\begin{equation}
m_1 < m_2 < m_3
\;.
\label{E207}
\end{equation}

For the transition amplitudes of neutrinos with momentum $p$
we have
\begin{equation}
\arraycolsep=0cm
\begin{array}{rcl} \displaystyle
\cal{A}_{\nu_{\ell}\to\nu_{\ell'}}
\null & \null = \null & \null \displaystyle
\sum_{i}
U_{\ell'i}
\,
{\mathrm{e}}^{ - i E_i t }
\,
U_{\ell i}^{*}
\\ \displaystyle
\null & \null = \null & \null \displaystyle
{\mathrm{e}}^{ - i E_1 t }
\left\{
\sum_{i=2}^{3}
U_{\ell'i}
\left[
\exp
\left(
- i
{\displaystyle
\Delta m^2_{i1} L
\over\displaystyle
2 p
}
\right)
- 1
\right]
U_{\ell i}^{*}
+
\delta_{\ell'\ell}
\right\}
\;.
\end{array}
\label{E208}
\end{equation}
Here
$ \displaystyle
E_i
=
\sqrt{ p^2 + m_i^2 }
\simeq
p
+
{\displaystyle
m_i^2
\over\displaystyle
2 p
}
$,
$ \Delta m^2_{i1} \equiv m^2_i - m^2_1 $
and
$ L \simeq t $ is the distance between
the neutrino source and detector.

Strong indications
in favor of neutrino mixing were obtained
in solar neutrino experiments
\cite{SOLAREXP}.
{}From the results of these experiments it follows
that
$ \Delta m^2_{21} \simeq 10^{-5} \, \mathrm{eV}^2 $
(if the resonant MSW mechanism is assumed)
or
$ \Delta m^2_{21} \simeq 10^{-10} \, \mathrm{eV}^2 $
(if vacuum oscillations are assumed).
If we accept for
$ \Delta m^2_{21} $
a value that can accommodate the solar neutrino data, we have,
for the experiments with terrestrial and atmospheric
neutrinos,
\begin{equation}
{\displaystyle
\Delta m^2_{21} \, L
\over\displaystyle
2 p
}
\ll
1
\;.
\label{E209}
\end{equation}

{}From
Eqs.(\ref{E208}) and (\ref{E209}),
for the probability of
$ \nu_{\ell} \to \nu_{\ell'} $
($ \bar\nu_{\ell} \to \bar\nu_{\ell'} $)
transitions with $\ell'\not=\ell$
we obtain the following expression
\cite{BFP92}:
\begin{equation}
P_{\nu_{\ell}\to\nu_{\ell'}}
=
P_{\bar\nu_{\ell}\to\bar\nu_{\ell'}}
=
{1\over2}
\,
A_{\nu_{\ell};\nu_{\ell'}}
\left(
1
-
\cos
{\displaystyle
\Delta m^2 \, L
\over\displaystyle
2 \, p
}
\right)
\;.
\label{E210}
\end{equation}
Here
$
\Delta m^2
\equiv
\Delta m^2_{31}
$
and
the amplitude of oscillations
$ A_{\nu_{\ell};\nu_{\ell'}} $
is given by
\begin{equation}
A_{\nu_{\ell};\nu_{\ell'}}
=
A_{\nu_{\ell'};\nu_{\ell}}
=
4
\left| U_{\ell3} \right|^2
\left| U_{\ell'3} \right|^2
\;.
\label{E211}
\end{equation}

For the probability of
$\nu_{\ell}$ ($\bar\nu_{\ell}$)
to survive we have
\begin{equation}
P_{\nu_{\ell}\to\nu_{\ell}}
=
P_{\bar\nu_{\ell}\to\bar\nu_{\ell}}
=
1
-
\sum_{\ell'\not=\ell}
P_{\nu_{\ell}\to\nu_{\ell'}}
=
1
-
{1\over2}
\,
B_{\nu_{\ell};\nu_{\ell}}
\left(
1
-
\cos
{\displaystyle
\Delta m^2 \, L
\over\displaystyle
2 \, p
}
\right)
\;,
\label{E212}
\end{equation}
where the oscillation amplitude
$ B_{\nu_{\ell};\nu_{\ell}} $
is given by
\begin{equation}
B_{\nu_{\ell};\nu_{\ell}}
=
\sum_{\ell'\not=\ell}
A_{\nu_{\ell};\nu_{\ell'}}
=
4
\left| U_{\ell3} \right|^2
\left(
1
-
\left| U_{\ell3} \right|^2
\right)
\;.
\label{E213}
\end{equation}

Thus,
in the model under consideration
all channels of neutrino oscillations
are characterized by the
same oscillation length
$
L_{\mathrm{osc}}
=
4 \pi p / \Delta m^2
$.
The amplitudes of oscillations are determined by
two parameters:
$ \left| U_{e3} \right|^2 $
and
$ \left| U_{\mu3} \right|^2 $.
(Due to the unitarity of the
mixing matrix
$
\left| U_{\tau3} \right|^2
=
1
-
\left| U_{e3} \right|^2
-
\left| U_{\mu3} \right|^2
$.)

\begin{figure}[t]
\begin{minipage}[t]{0.49\linewidth}
\begin{center}
\mbox{\epsfig{file=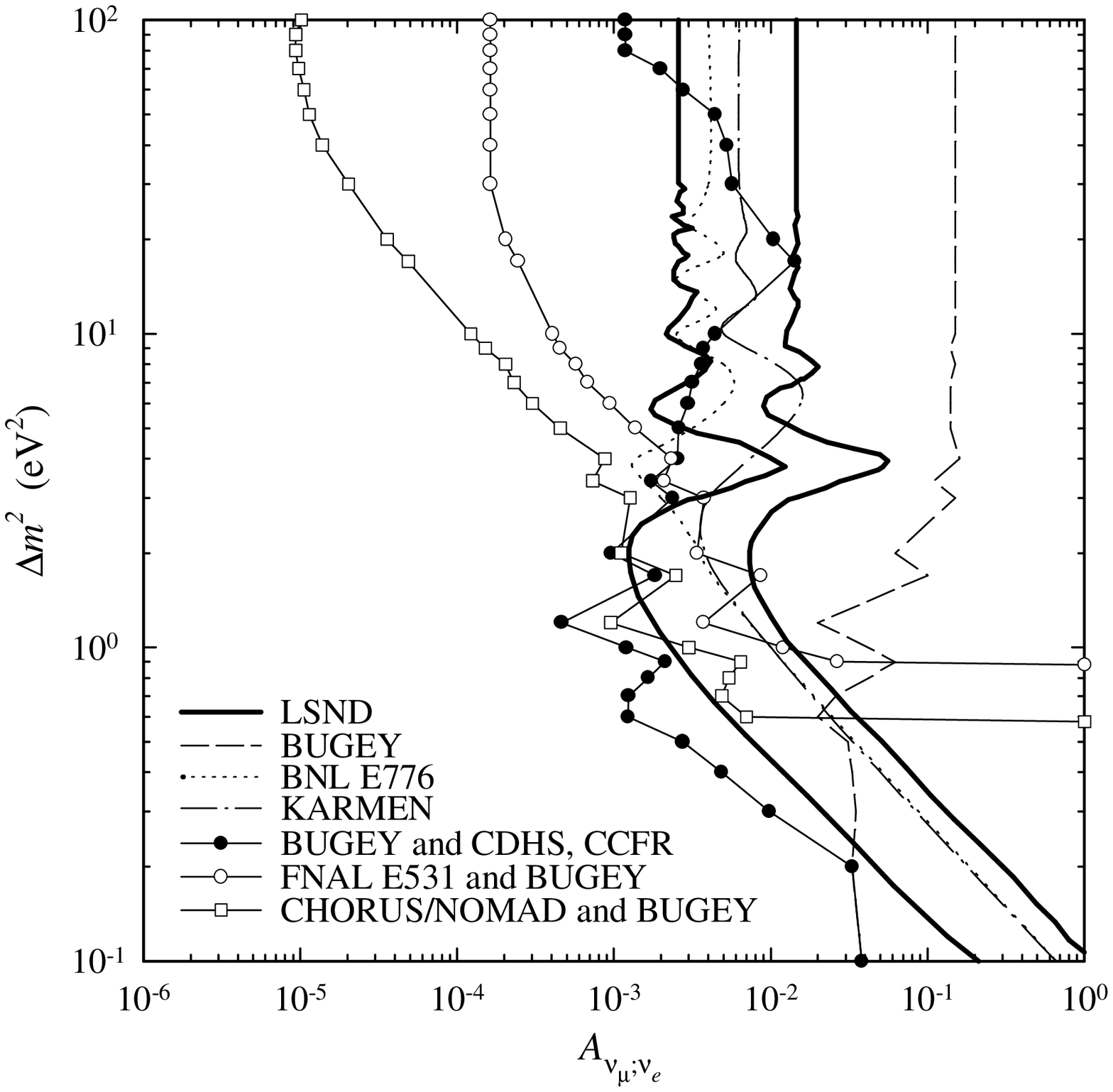,width=\linewidth}}
\\
\vspace{0.5cm}
Figure \ref{FIG1}
\end{center}
\end{minipage}
\null
\vspace{-0.3cm}
\null
\refstepcounter{figs}
\label{FIG1}
\hfill
\begin{minipage}[t]{0.49\linewidth}
\begin{center}
\mbox{\epsfig{file=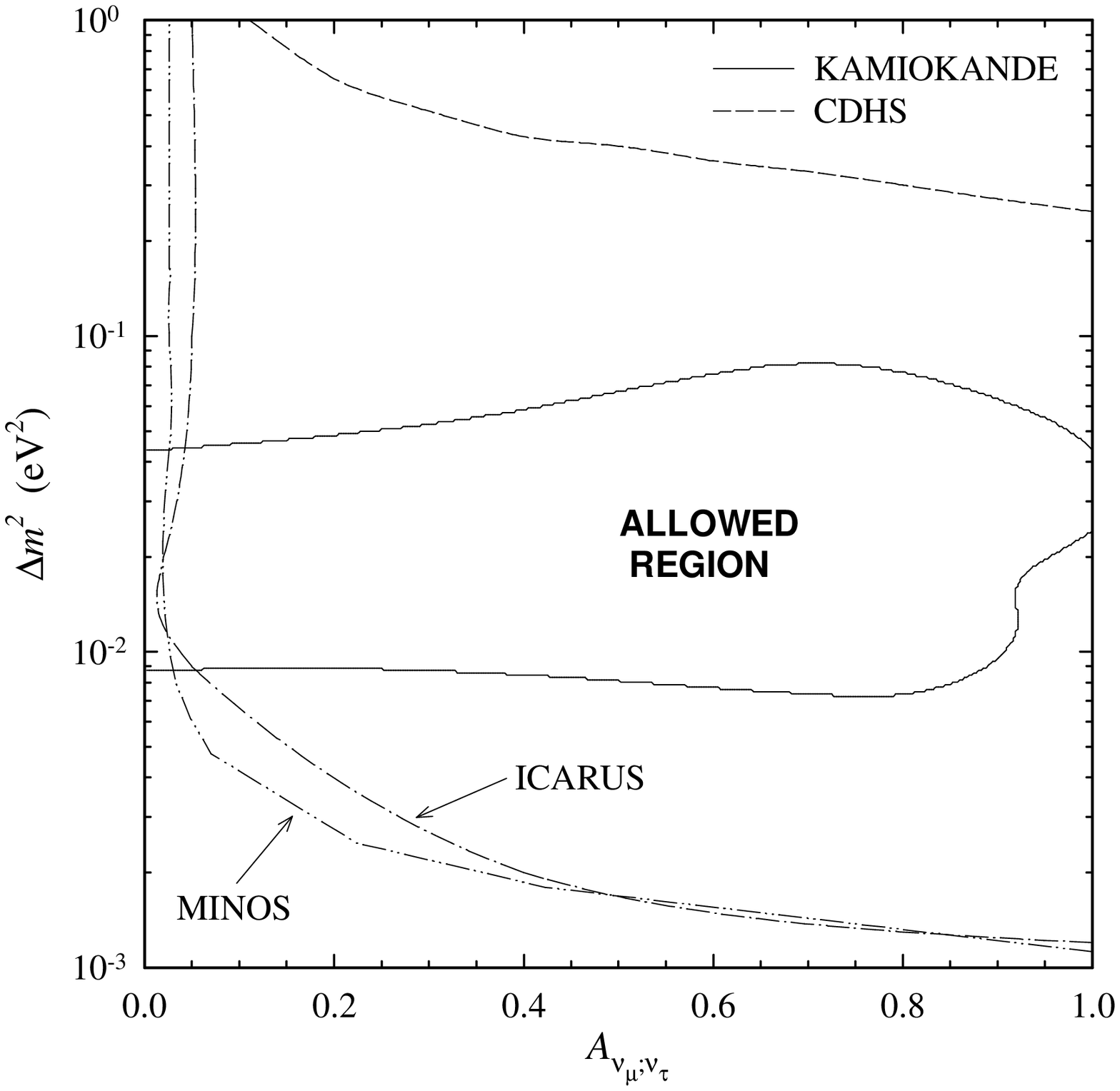,width=\linewidth}}
\\
\vspace{0.5cm}
Figure \ref{FIG2}
\end{center}
\end{minipage}
\null
\vspace{-0.3cm}
\null
\refstepcounter{figs}
\label{FIG2}
\end{figure}

We will discuss first
$ \nu_\mu \leftrightarrows \nu_e $
oscillations.
Experiments to search for
$ \nu_\mu \leftrightarrows \nu_e $
oscillations
are presently under way in Los Alamos
(LSND \cite{LSND})
and in Rutherford Laboratory
(KARMEN \cite{KARMEN}).
If there is a hierarchy of couplings among generations in the
lepton sector,
from the results of reactor and accelerator disappearance
experiments and also from the results of the
FNAL E531 experiment
\cite{E531}
which searched for
$ \nu_\mu \leftrightarrows \nu_\tau $
oscillations
it is possible to obtain
restrictions on the allowed values of the parameters of
$ \nu_\mu \leftrightarrows \nu_e $
oscillations
which are more severe than those obtained
from direct
$ \nu_\mu \to \nu_e $
appearance experiments
\cite{BBGK95}.
In Fig.\ref{FIG1}
we have presented the upper bounds for the amplitude
$ A_{\nu_{\mu};\nu_{e}} $
in the range
$ 10^{-1} \, \mathrm{eV}^2 \le \Delta m^2 \le 10^{2} \, \mathrm{eV}^2 $
obtained from exclusion plots of reactor and accelerator experiments
with the assumption that there is a hierarchy of couplings.
The curve passing through the filled circles
was obtained from the exclusion plots of the Bugey reactor experiment
\cite{BUGEY}
and
the $\nu_\mu$ disappearance
CDHS
\cite{CDHS}
and CCFR
\cite{CCFR} accelerator
experiments.
The curve passing through
the open circles was obtained from the results of the Bugey experiment
and the results of the FNAL E531 experiment,
which searched for
$ \nu_\mu \to \nu_\tau $
transitions.
We have also plotted the exclusion curves obtained in
the KARMEN (dash-dotted line) and BNL E776
\cite{E776}
(dotted line) experiments searching for
$ \nu_\mu \to \nu_e $
transitions.
Finally,
taking into account that
$
A_{\nu_{\mu};\nu_{e}}
\le
B_{\nu_{e};\nu_{e}}
$,
we also plotted in Fig.\ref{FIG1}
 the exclusion plot for
$ B_{\nu_{e};\nu_{e}} $
found in the Bugey experiment (dashed line).

The CHORUS
\cite{CHORUS}
and NOMAD
\cite{NOMAD}
experiments searching for
$ \nu_\mu \to \nu_\tau $
transitions are under way at CERN.
In Fig.\ref{FIG1} we have also plotted the range of
$ A_{\nu_{\mu};\nu_{e}} $
that could be
explored when the projected sensitivity of these experiments
is reached.

It was reported recently
\cite{LSND}
that a excess of 9
electron events with a expected background of
$ 2.1 \pm 0.3 $ events
was observed in the LSND experiment.
Under the assumption that these events are due to
$ \bar\nu_\mu \leftrightarrows \bar\nu_e $
oscillations,
the authors found the allowed
region in the plane of the parameters
$ A_{\nu_{\mu};\nu_{e}} $--$ \Delta m^2 $.
We have shown this region between the two
solid curves in Fig.\ref{FIG1}.
{}From Fig.\ref{FIG1} it is seen
that the LSND allowed region is within the region that is
forbidden by all the other existing data  on the
terrestrial  neutrino oscillation experiments.
Thus,
if the LSND result will be confirmed by future experiments,
it will mean that there is no natural hierarchy
of couplings in the lepton sector.

In fact,
in Ref.\cite{BBGK95}
we have shown that the observation of
$ \nu_\mu \leftrightarrows \nu_e $
oscillations
with an amplitude larger than the limit given in Fig.\ref{FIG1}
can be in agreement with all existing data on
neutrino oscillations
only if
$ \left| U_{\mu3} \right|^2 $
is large and
$ \left| U_{e3} \right|^2 $
is small
(under the assumption that
the model with mixing of three massive neutrino fields
and a neutrino mass hierarchy turns out to be correct).
This result would mean that the neutrino mixing matrix is
basically different from the Cabibbo-Kobayashi-Maskawa mixing matrix of
quarks and,
in this case,
we would confront
with a very unusual situation
in which $\nu_\mu$
(and not $\nu_\tau$ as in the case of hierarchy)
is the heaviest neutrino.
Let us notice also that in the case of
large
$ \left| U_{\mu3} \right|^2 $
and small
$ \left| U_{e3} \right|^2 $
the results of the experiments to
search for
$ \nu_\mu \leftrightarrows \nu_\tau $
oscillations will not allow to make
any prediction about the outcome of the experiments to
search for
$ \nu_\mu \leftrightarrows \nu_e $
oscillations and vice versa.

So far we have considered the experiments on the search for oscillations
of reactor and accelerator neutrinos which could  reveal
the effects of neutrino mixing if
$ \Delta m^2 $
is in the range
$ 10^{-1} \, \mathrm{eV}^2 \le \Delta m^2 \le 10^{2} \, \mathrm{eV}^2 $.
To further expand the region of $\Delta m^2$ to be explored,
we have considered also
\cite{BGK95}
the atmospheric neutrino experiments and the long-baseline
neutrino experiments under preparation, which are
sensitive to much smaller values of the parameter
$\Delta m^2$
($ \Delta m^2 \gtrsim 10^{-3} \, \mathrm{eV}^2 $).

As it is well known, the so-called atmospheric neutrino anomaly
was observed in three underground neutrino
experiments
(Kamiokande \cite{KAM94},
IMB \cite{IMB}
and
Soudan 2 \cite{SOUDAN}).
For the ratio
\begin{equation}
R
\equiv
{\displaystyle
N_\mu^{\mathrm{exp}} / N_e^{\mathrm{exp}}
\over\displaystyle
N_\mu^{\mathrm{MC}} / N_e^{\mathrm{MC}}
}
\label{E214}
\end{equation}
the following value was obtained
in the Kamiokande experiment:
$ R = 0.60^{+0.06}_{-0.05} \pm 0.05 $.
Here
$N_e^{\mathrm{exp}}$ and $N_\mu^{\mathrm{exp}}$
are the numbers of $e$-like and $\mu$-like
events and
$N_e^{\mathrm{MC}}$ and $N_\mu^{\mathrm{MC}}$
are the numbers of
$e$-like and $\mu$-like
events predicted by the MC simulation under the assumption that there
are no neutrino oscillations.
The IMB and Soudan 2 experiments
obtained values of $R$
compatible with that found by the Kamiokande experiment.

If there is a real anomaly for atmospheric neutrinos,
the most natural
explanation of it is neutrino oscillations.
{}From the analysis of the Kamiokande data in the simplest case
of oscillations between two types of neutrinos the following
allowed ranges for the parameters
$ \Delta m^2 $
and
$ \sin^2 2\vartheta $
($\vartheta$ is the mixing angle)
have been found
\cite{KAM94}:
\begin{equation}
5 \times 10^{-3}
\le
\Delta m^2
\le
3 \times 10^{-2} \, \mbox{eV}^2
\quad \quad
0.7
\le
\sin^2 2\vartheta
\le
1
\label{E072}
\end{equation}
in the case of
$ \nu_{\mu} \leftrightarrows \nu_{\tau} $
oscillations
and
\begin{equation}
7 \times 10^{-3}
\le
\Delta m^2
\le
8 \times 10^{-2} \, \mbox{eV}^2
\quad \quad
0.6
\le
\sin^2 2\vartheta
\le
1
\label{E071}
\end{equation}
in the case of
$ \nu_{\mu} \leftrightarrows \nu_{e} $
oscillations.

At present a wide range of long baseline experiments
with neutrinos from reactors and accelerators
aimed to search for
neutrino oscillations
in different channels
($ \nu_{e} \leftrightarrows \nu_{e} $
with reactor neutrinos
and
$ \nu_{\mu} \leftrightarrows \nu_{\mu} $,
$ \nu_{\mu} \leftrightarrows \nu_{e} $,
$ \nu_{\mu} \leftrightarrows \nu_{\tau} $
with accelerator neutrinos)
in the
``atmospheric neutrino range of $ \Delta m^2 $''
are under preparation
\cite{CHOOZ,KEK,E889,MINOS,ICARUS}.
We have analyzed the atmospheric neutrino data in the
framework of the model with mixing of three massive
neutrinos discussed above
\cite{BGK95}.
We have found that
the atmospheric neutrino data are well described by this model
and we have obtained the allowed regions of the oscillation
parameters for different oscillation channels that will be
investigated in future long baseline neutrino experiments.
We present here only the Kamiokande-allowed region
in the channel
$ \nu_{\mu} \leftrightarrows \nu_{\tau} $
(Fig.\ref{FIG2}).
In Fig.\ref{FIG2} we have shown also the region of sensitivity
of the future
MINOS \cite{MINOS} (Fermilab--Soudan)
and
ICARUS \cite{ICARUS} (CERN--Gran Sasso)
experiments,
in both of which the detector will be displaced
at a distance of about 730 Km from the source.
As it is seen from Fig.\ref{FIG2},
a large part of the region in the
$ A_{\nu_\mu;\nu_\tau} $--$ \Delta m^2 $
plane
that is allowed by Kamiokande data
will be investigated in long baseline experiments.

The problem of neutrino masses and mixing is under very active
investigation and
many experiments
that are sensitive to a very wide range of neutrino
mixing parameters are under way at present.
{}From the solar neutrino experiments rather weighty
indications in favor of nonzero neutrino
masses and mixing have been found.
Undoubtedly, future experiments  will lead to a significant progress in
the investigation of the fundamental properties of neutrinos.

\end{document}